\documentclass[apl,apl,reprint,aps,showpacs,showkeys,twocolumn]{revtex4-1}

\pdfoutput=1

\usepackage[utf8]{inputenc}
\usepackage[T1]{fontenc}
\usepackage{hyperref}
\usepackage{color}
\usepackage{graphicx}
\usepackage{units}
\usepackage{amsmath}

\hypersetup{colorlinks = true, citecolor = blue, breaklinks = true}

\begin{document}

\title{Interplay between strain, defect charge state and functionality in complex oxides}
\date{\today}
\author{Ulrich Aschauer}
\affiliation{Materials Theory, ETH Zurich, Wolfgang-Pauli-Strasse 27, CH-8093 Z\"urich, Switzerland}
\affiliation{Department of Chemistry and Biochemistry, University of Bern, Freiestrasse 3, CH-3012 Bern, Switzerland}
\author{Nicola A. Spaldin}
\affiliation{Materials Theory, ETH Zurich, Wolfgang-Pauli-Strasse 27, CH-8093 Z\"urich, Switzerland}

\begin{abstract}
We use first-principles calculations to investigate the interplay between strain and the charge state of point defect impurities in complex oxides. Our work is motivated by recent interest in using defects as active elements to provide novel functionality in coherent epitaxial films. Using oxygen vacancies as model point defects, and CaMnO$_3$ and MnO as model materials, we calculate the changes in internal strain caused by changing the charge state of the vacancies, and conversely the effect of strain on charge-state stability. Our results show that the charge state is a degree of freedom that can be used to control the interaction of defects with strain and hence the concentration and location of defects in epitaxial films. We propose the use of field-effect gating to reversibly change the charge state of defects and hence the internal strain and corresponding strain-induced functionalities. 
\end{abstract}

\maketitle

Defects in oxide thin films are increasingly being considered as active elements that can lead to added functionality \cite{Kalinin:2012fs,Kalinin:2013kd,Biskup:2014eo,Farokhipoor:2015kq,Becher:2015df}. One promising route to engineering new material behaviors is to exploit the simultaneous coupling between strain and structural or electronic properties, and strain and defect formation energies. This concept is well established in bulk ceramics, where defects are known to cause isotropic or anisotropic changes in volume \cite{Grande:2012ib}. In the case of oxygen vacancies, the extra electrons remaining after removal of a neutral oxygen atom reduce the surrounding transition metal ions and increase their radii, leading to a local {\it chemical expansion} \cite{Adler:2001ww} of the lattice; the opposite effect is seen for cation vacancies. Conversely, volume changes achieved through biaxial strain affect the defect formation energy and hence the concentration of defects \cite{Aschauer:2013hf}. In the context of thin films, the discussion to date has focussed largely on neutral defects, for which these simple volumetric arguments are often straightforwardly applicable. Defects also form, however, in \textit{charged} configurations, and the amount of reduction or oxidation of the surrounding transition-metal ions of course depends on the defect charge state.

In this work we extend the discussion of the interplay between strain, oxygen vacancy formation energy and functionality to the case of charged vacancies. We use density functional theory to constrain the charge state of the defect and to selectively suppress the transition-metal reduction and associated lattice expansion that accompanies the formation of neutral oxygen vacancies. For the model material CaMnO$_3$, we find that the response of the lattice to oxygen vacancy formation in the absence of the transition-metal reduction is a contraction, with a lattice expansion only observed when transition metal reduction occurs. Our findings suggest that dynamical control of the charge state via electrostatic gating could result in drastic but controllable changes in the structure and corresponding properties of the thin film. To explore this behavior, we analyze how the defect charge state is affected by the band alignment between the substrate and the film in a thin-film geometry. Finally, we show that in the case of MnO, where the residual charge left by a neutral oxygen vacancy is localized in an F-center rather than on the surrounding transition metal ions, the ionic radius change at the transition metal is minimal and intriguingly even neutral vacancy formation results in a lattice contraction.

Our DFT calculations were performed with the VASP code \cite{Kresse:1993ty,Kresse:1994us,Kresse:1996vk,Kresse:1996vf} with the same computational parameters as previously published for CaMnO$_3$ \cite{Aschauer:2013hf} and MnO \cite{Aschauer:2015dd}. Formation energies of charged oxygen vacancies \cite{Freysoldt:2014ej} were computed as
\begin{equation}
\Delta E_\mathrm{form} = E_\mathrm{def} - E_\mathrm{stoi} - \mu_\mathrm{O} + q E_\mathrm{fermi} + E_\mathrm{corr}
\end{equation}
where $E_\mathrm{def}$ and $E_\mathrm{stoi}$ are the DFT total energies of the defective and stoichiometric supercells respectively and $\mu_\mathrm{O}$ is the chemical potential that accounts for the removal of an oxygen atom. For charged cells, $q$ electrons are exchanged with the electron reservoir at the Fermi energy $E_\mathrm{fermi}$ (referenced to the valence band edge) and $E_\mathrm{corr}$ is a corrective term to align the potentials of the neutral stoichiometric and the charged defective cell. $E_\mathrm{corr}$ was taken to be the potential energy difference between the neutral and charged cells at the location of an ion far from the defect \cite{Lany:2008gk} and we verified that a more involved approach based on a model charge density \cite{Freysoldt:2009ih,Freysoldt:2010gx} yields an equivalent result within the accuracy of our calculations.

\begin{figure}
\includegraphics[width=\columnwidth]{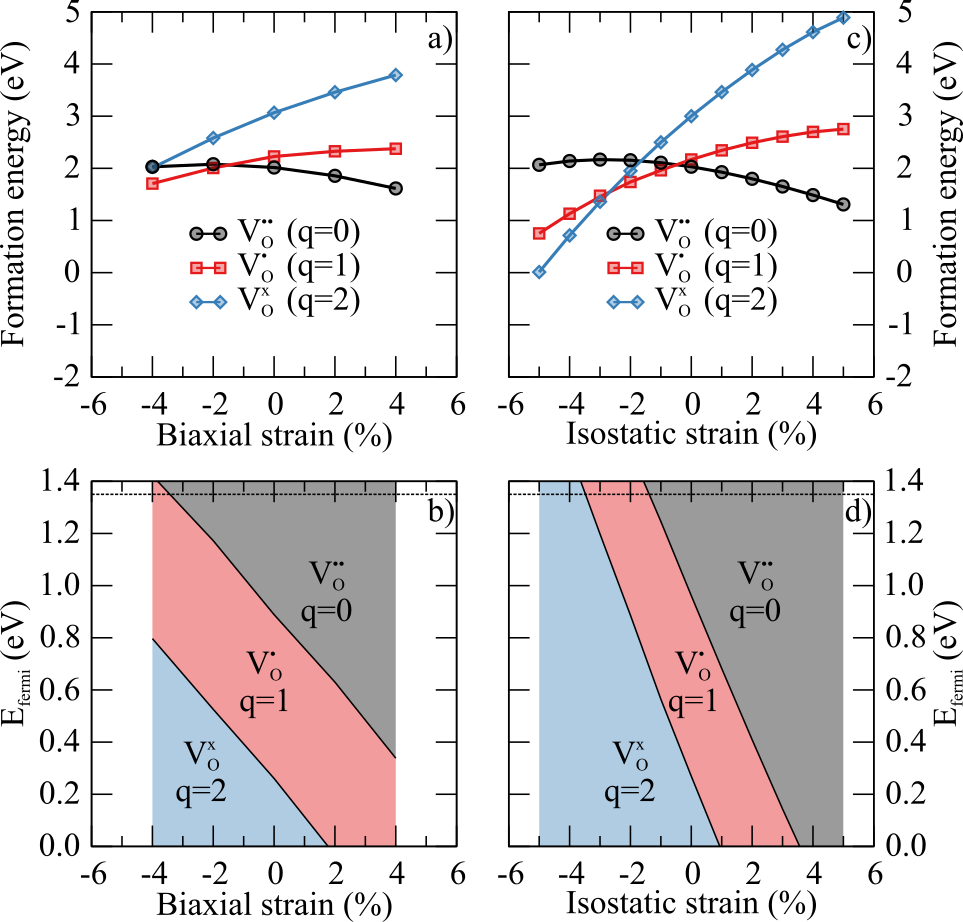}
\caption{\label{fig:CMO}Formation energy of neutral, singly and doubly charged oxygen vacancies in CaMnO$_3$ as a function of a) biaxial strain and c) isostatic strain. Formation energies were computed for $\mu_\mathrm{O}$ = -5 eV and E$_\mathrm{fermi}$ = 1.1 eV. Panels b) and d) show the evolution of the transition levels with applied biaxial and isostatic strain respectively.}
\end{figure}

In Fig. \ref{fig:CMO}a) we show our calculated formation energies for oxygen vacancies of three different charge states as a function of biaxial strain in the perovskite oxide CaMnO$_3$. This geometry is relevant for the case of coherent heteroepitaxial thin films, which are central to many areas of technology and research. Under biaxial strain, the length of only two out of the three axes is imposed, with the third free to relax according to the Poisson ratio of the material. Despite this free axis, tensile strain leads to a increased unit-cell volume in CaMnO$_3$, while compressive strain reduces it \cite{Aschauer:2013hf}. For clarity, we report the formation energy at the energetically more favorable in-plane vacancy position \cite{Aschauer:2013hf}, having verified that the formation energy at the out-of-plane position follows a similar trend as a function of the charge state. We chose an oxygen chemical potential $\mu_\mathrm{O}$ = -5 eV, corresponding to the situation in air \cite{Aschauer:2013hf} and a Fermi energy of 1.1 eV, which is close to the theoretical band gap of 1.35 eV. We see that the formation energy of the neutral oxygen vacancy (V$_\mathrm{O}^{\bullet\bullet}$, q=0) decreases under volume expansion, as expected from the chemical pressure effect discussed above. In contrast, the singly (V$_\mathrm{O}^{\bullet}$, q=1) and doubly (V$_\mathrm{O}^x$, q=2) charged vacancies have lower formation energies under compressive strain, with the doubly charged vacancy showing the strongest strain dependence. As a result, under the chosen conditions of chemical potential and Fermi energy, neutral vacancies form preferentially at ambient or tensile strains, singly charged vacancies for compressive strains larger than 1.5\%, while doubly charged vacancies would form at compressive strains larger than the 4\% considered here.

These trends are consistent with the discussion above: Formation of a neutral V$_\mathrm{O}^{\bullet\bullet}$ is accompanied by the reduction of two Mn sites adjacent to the vacancy (Mn$^{4+} \rightarrow$ Mn$^{3+}$), which increases their ionic size by $\sim$0.115 \AA\ (the Shannon ionic radius of 6-coordinated Mn$^{4+}$ = 0.530 \AA\ and that of Mn$^{3+}$ = 0.645 \AA \cite{Shannon:1976vx}). Consequently neutral oxygen vacancies are favored when the volume is expanded under tensile strain and the size increase is more easily accommodated. The reduction occurs on only one Mn for V$_\mathrm{O}^{\bullet}$ and not at all for V$_\mathrm{O}^x$. For V$_\mathrm{O}^{\bullet}$ the tendency for volume increase is thus weaker and is counteracted by the creation of empty space at the vacancy site; for V$_\mathrm{O}^x$ there is no expansion mechanism and the vacancies tend to strongly reduce the volume.  We emphasize then that the usual chemical expansion associated with oxygen-vacancy formation in transition-metal oxides occurs only for neutral oxygen vacancies and is a result of the reduction of the adjacent transition-metal ions. At least in the case of CaMnO$_3$, the intrinsic effect of the oxygen vacancies on the lattice without an accompanying reduction reaction is in fact a contraction. The same contraction for charged vacancies was also reported for SrTiO$_3$ \cite{Aidhy:2015js}, where the doubly-charged oxygen vacancy is stable for all values of the Fermi energy \cite{Janotti:2014bn}.

The stability of oxygen vacancies in different charge states strongly depends on the position of the Fermi energy. We illustrate this in Fig. \ref{fig:CMO}b), where we show the ranges of stability of the different charge states (separated by the transition levels) as a function of biaxial strain and the position of the Fermi level. We see that for Fermi energies close to the conduction band edge of 1.35 eV (dashed horizontal line in Fig. \ref{fig:CMO}b), which corresponds to a physically relevant situation, the neutral defect is stable across a wide strain range, with compressive strains close to 4\% being required to stabilise the V$_\mathrm{O}^{\bullet}$ defect. Conversely, for this $E_\mathrm{fermi}$, strong compressive strain applied to a system containing neutral oxygen vacancies would cause spontaneous charging of the vacancies and a corresponding strain compensation.

In Fig. \ref{fig:CMO}c) we show the corresponding calculated formation energies as a function of isostatic, rather than biaxial, strain.  Since we now control the length of all three axes, we expect the strain dependence of the vacancy formation energy to be stronger than in the biaxial case. Indeed our calculations confirm this expectation: Comparing the biaxial situation in Fig. \ref{fig:CMO}a) to the isostatic one in Fig. \ref{fig:CMO}c), we see the same general trends of neutral V$_\mathrm{O}^{\bullet\bullet}$ favoured under tensile strain, and charged vacancies favoured under compressive strain. We notice, however, that the strain dependence is much stronger, so that under the same conditions of $\mu_\mathrm{O}$ and E$_\mathrm{fermi}$ the charged defects become stable at smaller compressive strains of around 0.5\% and 3\% for the singly charged (V$_\mathrm{O}^{\bullet}$) and the doubly charged  (V$_\mathrm{O}^x$) vacancy respectively. The ranges of stability as a function of isostatic strain and the Fermi energy are again shown in Fig. \ref{fig:CMO}d), showing that compared to the biaxial case the neutral defect is stable over a smaller range of strains, compressive isostatic strain quickly promoting the formation of charged defects. We also note that the slope of the transition levels as a function of strain is much more pronounced than for biaxial strain, indicating again the stronger coupling between strain and defect formation.

\begin{figure}
\includegraphics[width=\columnwidth]{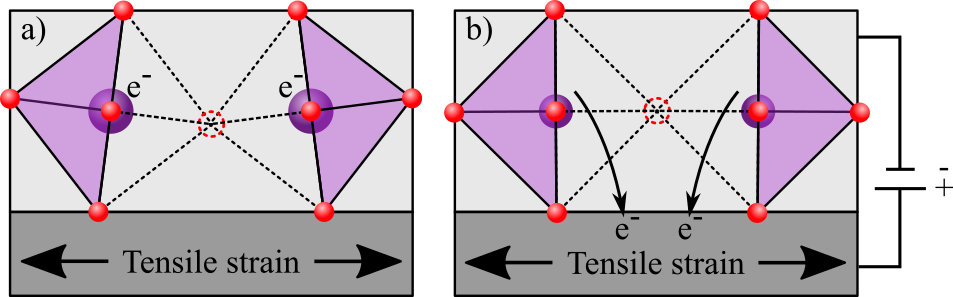}
\caption{\label{fig:gating}a) Epitaxial film with a neutral oxygen vacancy to relax tensile strain. b) Upon gating the extra electrons are transferred to the substrate and the film changes its structure (bond lengths and angles) to accommodate the strain.}
\end{figure}

In practice, the Fermi energy of a thin film can be tuned dynamically using electrostatic gating. Injection of electrons using a field effect into a material that has formed neutral oxygen vacancies to relax tensile strain (Fig. \ref{fig:gating}a) would convert the defects from local expansion to local contraction centres, and the film would experience a renewed tensile strain. The film would then have to find other means to accommodate the strain, which we expect to happen through structural changes of the bond lengths (Fig. \ref{fig:gating}b) and angles \cite{Rondinelli:2011jk}. Gating of oxygen deficient films could thus be a way to gradually and reversibly change the structure of the interfacial region and allow activation of properties induced by tensile strain such as ferroelectricity or local changes in magnetic order \cite{Bhattacharjee:2009eb,Lee:2010kz,Gunter:2012ki,Iusan:2013fl}.

\begin{figure}[b]
\includegraphics[width=\columnwidth]{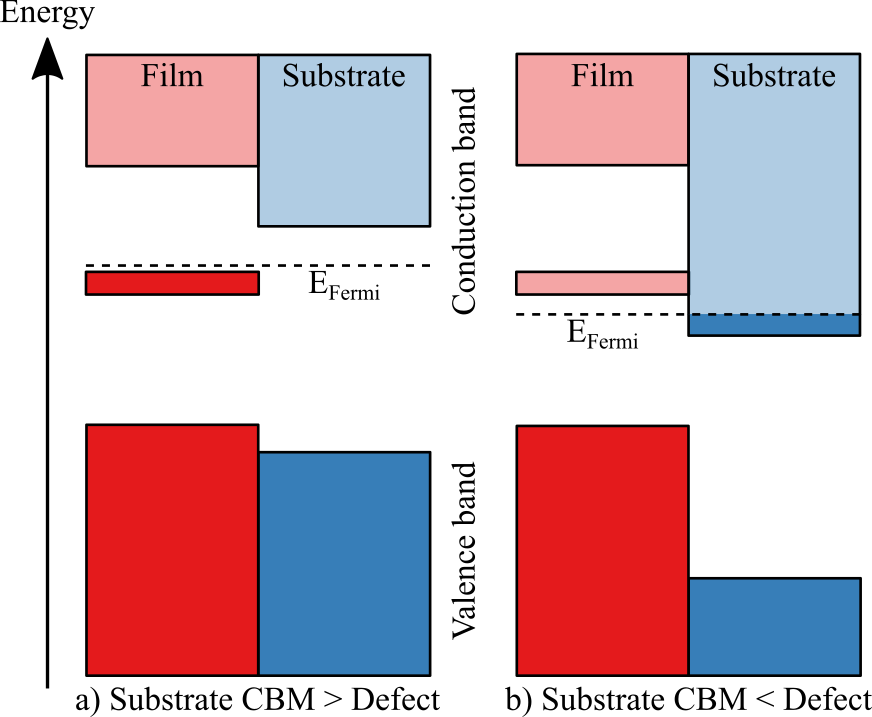}
\caption{\label{fig:substrate}Schematic illustration of the effect of the band alignment between film (red) and substrate (blue) on the charge state of the defect: a) filled defect state (V$_\mathrm{O}^{\bullet\bullet}$) when the substrate conduction-band edge is above the defect state and b) empty defect state (V$_\mathrm{O}^x$) when the substrate conduction-band edge is below the defect state.}
\end{figure}

We note, however, that for a thin film in contact with a substrate, the most favourable charge state is determined in part by the band alignment between the thin film and the substrate. In Figure \ref{fig:substrate} we show two schematic band alignments between substrates and strained films. In panel a) the conduction band edge of the substrate is higher in energy than the defect state and so electrons are most favourably accommodated in the defect state resulting in a neutral oxygen vacancy (V$_\mathrm{O}^{\bullet\bullet}$), which readily accommodates tensile strain. In b), where the substrate conduction-band edge is at lower energy than the defect state, the electrons will tend to transfer to the substrate conduction band leaving a charged V$_\mathrm{O}^x$ defect that is compatible with compressive strain. While for the band alignment shown in Fig.~\ref{fig:substrate}a)  tensile strain leads to more oxygen vacancies as we discussed above, for the band alignment shown in Fig. \ref{fig:substrate}b) the opposite is true and tensile strain should in fact suppress the formation of oxygen vacancies. The choice of substrate thus not only imposes the strain, but its band alignment with the film determines the type and charge state of defects that form to accommodate the strain.

\begin{figure}
\includegraphics[width=\columnwidth]{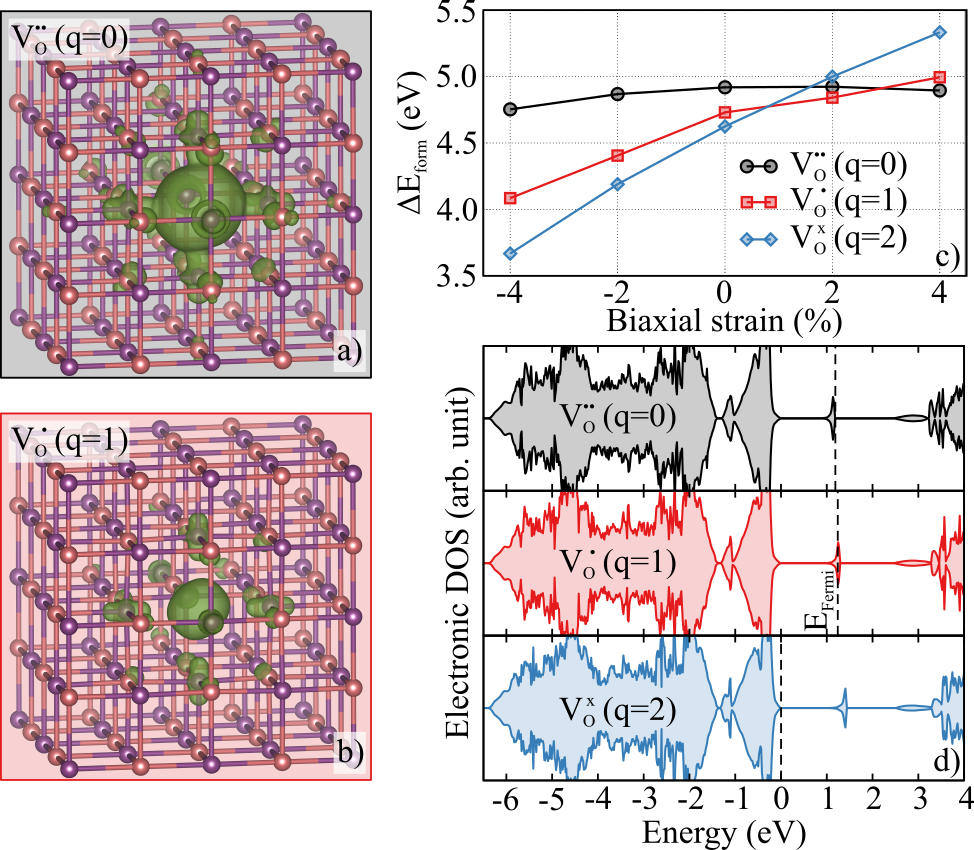}
\caption{\label{fig:MnO}Charge density associated with this defect state for a) the neutral (V$_\mathrm{O}^{\bullet\bullet}$) and b) the singly charged (V$_\mathrm{O}^{\bullet}$) defect respectively. Panel c) shows the formation energy of an oxygen vacancy in MnO in the neutral and charged states and d) the respective electronic densities of states, where the defect state about 1.5 eV above the valence-band edge is visible.}
\end{figure}

In the prototype compound CaMnO$_3$ discussed above, the additional vacancy electrons reduce the neighboring Mn$^{4+}$ to high-spin Mn$^{3+}$ increasing the ionic radius and leading to the predicted simple strain dependence. In this final section we analyze the more complicated situation found in MnO, in which the reduction effect and consequently the change in radius associated with neutral oxygen vacancy formation is much smaller. In MnO the extra charge introduced by neutral oxygen vacancy formation does not primarily localise on transition-metal sites but instead forms an F-center at the vacancy site \cite{Aschauer:2015dd}. We illustrate this in Figs. \ref{fig:MnO}a) and b), where we plot the defect-state charge densities for constrained neutral V$_\mathrm{O}^{\bullet\bullet}$ and singly charged V$_\mathrm{O}^{\bullet}$ vacancies. The characteristic F-center behavior, in which the defect charge is primarily localised on the vacancy site with only small tails on the surrounding Mn atoms, is clear. This charge localisation reflects the difficulty of reducing Mn$^\mathrm{2+}$ cations to Mn$^\mathrm{1+}$.

In Fig. \ref{fig:MnO}c) we plot the oxygen vacancy formation energies as a function of biaxial strain under the conditions $\mu_\mathrm{O}$=-4.58 eV, E$_\mathrm{Fermi}$=0.8 eV corresponding to ambient conditions \cite{Aschauer:2015dd} and a Fermi energy lower than the defect state, reflecting the typical p-type nature of MnO. We see that the neutral defect (V$_\mathrm{O}^{\bullet\bullet}$) has the highest formation energy of around 4.9 eV at 0\% strain. Removing electrons results in the more favourable singly V$_\mathrm{O}^{\bullet}$ and doubly V$_\mathrm{O}^x$ charged defects with formation energies of around 4.7 and 4.6 eV respectively at 0\% strain. We can understand this behavior straightforwardly from the electronic density of states (Fig. \ref{fig:MnO}d) where we see that removing electrons increasingly depopulates the defect state at around 1.5 eV above the valence-band edge.

Returning to the relative strain dependence of the individual formation energies (which is independent of the chosen chemical potential and Fermi energy) shown in Fig. \ref{fig:MnO}c), we see that the formation energy of the neutral (V$_\mathrm{O}^{\bullet\bullet}$) defect depends only weakly on the applied strain, reducing by 0.03 eV (not significant within our computational accuracy) and 0.17 eV under 4\% tensile and 4\% compressive strain respectively. The decrease in the formation energy of the neutral defect under compressive strain is opposite to the behavior of CaMnO$_3$, and confirms that vacancy-induced electrons that localize on defect sites and do not reduce the surrounding transition metal cations do not cause a lattice expansion. The reduction in formation energy with compressive strain for the charged defects (V$_\mathrm{O}^{\bullet}$ and V$_\mathrm{O}^x$), is even stronger, for the reasons that we discussed already above for CaMnO$_3$. This result shows that in certain situations, even neutral defects lower their formation energies under compressive strain. We anticipate similar behavior in metallic complex oxides, in which the excess charge is not well localised on specific cations.

In summary, we have shown that the magnitude and sign of the strain induced by defects in complex oxides depends sensitively on the charge state of the defect. Conversely, a change in the charge state of defects, which can be realized for example by using a field effect in a thin film geometry to add or remove carriers, will change the strain state and hence the properties of the material. The substrate in this case influences the response of the oxide both by imposing a strain via coherent heteroepitaxy and through the alignment of its band edge energies with the energy of the defect state. We hope that our calculations will motivate the development of thin film architectures to allow tunable and reversible changes in the strain state and consequently the functional properties of complex oxides.  

\section*{Acknowledgements}

This work was financially supported by the ETH Z\"urich and by the ERC Advanced Grant program, No. 291151. Computer resources were provided by the ETH Z\"urich (Euler cluster) and the Swiss Supercomputing Center (CSCS) under project s624.

\bibliography{references.bib}

\end{document}